\documentclass[sigconf]{acmart}
\AtBeginDocument{%
  }

\copyrightyear{2025}
\acmYear{2025}
\setcopyright{cc}
\setcctype{by}
\acmConference[MM '25]{Proceedings of the 33rd ACM International Conference on Multimedia}{October 27--31, 2025}{Dublin, Ireland}
\acmBooktitle{Proceedings of the 33rd ACM International Conference on Multimedia (MM '25), October 27--31, 2025, Dublin, Ireland}\acmDOI{10.1145/3746027.3758149}
\acmISBN{979-8-4007-2035-2/2025/10}

\author{Yili Jin}
\authornote{This work was done while Yili was a visiting student at Simon Fraser University.}
\affiliation{%
  \institution{McGill University}
  \city{Montreal}
  \country{Canada}}
\affiliation{%
  \institution{Simon Fraser University}
  \city{Burnaby}
  \country{Canada}}

\author{Xue Liu}
\affiliation{%
  \institution{Mohamed bin Zayed University of Artificial Intelligence}
  \city{Abu Dhabi}
  \country{UAE}}
\affiliation{%
  \institution{McGill University}
  \city{Montreal}
  \country{Canada}}

\author{Jiangchuan Liu}
\authornote{Jiangchuan Liu is the corresponding author.}
\affiliation{%
  \institution{Simon Fraser University}
  \city{Burnaby}
  \country{Canada}}

\usepackage{enumitem}

\begin{document}

\title[Generative AI for Multimedia Communication]{Generative AI for Multimedia Communication: Recent Advances, An Information-Theoretic Framework, and Future Opportunities}


\begin{abstract}
Recent breakthroughs in generative artificial intelligence (AI) are transforming multimedia communication. This paper systematically reviews key recent advancements across generative AI for multimedia communication, emphasizing transformative models like diffusion and transformers. However, conventional information-theoretic frameworks fail to address semantic fidelity, critical to human perception. We propose an innovative semantic information-theoretic framework, introducing semantic entropy, mutual information, channel capacity, and rate-distortion concepts specifically adapted to multimedia applications. This framework redefines multimedia communication from purely syntactic data transmission to semantic information conveyance. We further highlight future opportunities and critical research directions. We chart a path toward robust, efficient, and semantically meaningful multimedia communication systems by bridging generative AI innovations with information theory. This exploratory paper aims to inspire a semantic-first paradigm shift, offering a fresh perspective with significant implications for future multimedia research.
\end{abstract}





\maketitle
\section{Introduction}
Generative artificial intelligence (AI), including generative adversarial networks (GANs)~\cite{DBLP:conf/nips/GoodfellowPMXWOCB14}, transformers~\cite{DBLP:conf/nips/VaswaniSPUJGKP17}, and diffusion models~\cite{DBLP:conf/nips/HoJA20}, have rapidly advanced capabilities in synthesizing realistic multimedia content. This has revolutionized applications from video conferencing and streaming to augmented reality (AR) and virtual reality (VR), by enabling high-fidelity reconstruction of multimedia from minimal data. However, traditional multimedia communication frameworks, grounded in Shannon’s classical information theory, prioritize syntactic correctness, measured in bits and pixels, over semantic fidelity, the preservation of meaningful content perceived by humans.

In multimedia contexts, semantic fidelity often outweighs precise pixel-level accuracy. For instance, minor pixel distortions may be visually imperceptible or irrelevant if the intended semantic content, such as identifiable objects or spoken words, is accurately preserved. Classical information theory, by ignoring such semantic nuances, misses substantial opportunities for compression, error concealment, and quality enhancement.

Addressing this gap, our paper presents a new semantic-aware information-theoretic framework. We redefine key classical concepts, entropy, mutual information, and channel capacity, in semantic terms, thereby enabling multimedia communication systems to optimize for human-perceived quality and meaning rather than mere data fidelity. We outline recent advances in generative AI relevant to multimedia, demonstrating how generative models already implicitly leverage semantic priors to produce perceptually superior outputs. Looking ahead, this semantic perspective opens up transformative opportunities for designing communication systems that are more resilient, efficient, and adaptive, especially in bandwidth-limited, real-time, and immersive environments. It also lays the groundwork for future research on cross-modal compression, personalized generation, and AI-augmented communication protocols that prioritize meaning over data volume.


\section{Recent Advances}

\subsection{Multimedia Generation}

Recent generative AI techniques can produce highly realistic multimedia content, opening possibilities for content creation and more efficient communication.

\subsubsection{Image Generation}
Image generation has seen rapid progress due to deep generative models. GAN-based approaches like StyleGAN~\cite{DBLP:conf/cvpr/KarrasLAHLA20} achieved photorealistic image generation, and more recently diffusion models~\cite{DBLP:conf/icml/NicholDRSMMSC22} have taken the lead in quality. These diffusion models, trained on massive image datasets, can generate high-resolution images from text prompts or other inputs, vastly outperforming earlier GAN approaches in diversity and realism. Such models have been leveraged for image-based communication as well, for instance, generative codecs use an encoder-decoder to compress images into latent codes and then reconstruct high-quality images at the receiver.

\subsubsection{Video Generation}
Video generation is substantially more challenging than image generation due to the temporal dimension and coherence requirements. Early GAN-based video generators (e.g., MoCoGAN~\cite{DBLP:conf/cvpr/Tulyakov0YK18}) could produce short clips at low resolution. In the last few years, researchers have made major strides using improved GANs, autoregressive Transformers, and especially diffusion models. For example, Make-A-Video~\cite{DBLP:conf/iclr/SingerPH00ZHYAG23} is a diffusion model for text-conditioned video generation, which demonstrated that leveraging massive image pre-training and then unsupervised video fine-tuning can produce coherent, high-fidelity video clips from text prompts. Google’s Imagen~\cite{DBLP:conf/nips/HoSGC0F22} similarly introduced cascaded diffusion models to generate high-resolution videos from text, maintaining temporal consistency through explicit frame interpolation steps. Transformers have also been applied: e.g., Video Transformer models~\cite{DBLP:conf/eccv/GeHYYPJHP22} quantize frames with VQ-VAE and then use transformer decoding to generate long videos by modeling temporal dependencies. These techniques have key applications in communication systems. Generative models can predict and interpolate frames for video streaming, filling missing frames or reducing frame rates to save bandwidth. In real-time communication, neural video compression uses generative models to synthesize frames from compact signals, reducing data rates for video conferencing. This approach offers high-quality video reconstruction under low bandwidth by transmitting semantic descriptions or low-res signals.

\subsubsection{Audio Generation}
Generative AI is also transforming audio and speech, with applications in voice communication, streaming, and content creation. Neural text-to-speech models now routinely produce human-like speech from text. For instance, autoregressive models~\cite{DBLP:conf/ssw/OordDZSVGKSK16} and later GAN-based models~\cite{DBLP:conf/nips/KumarKBGTSBBC19} were early breakthroughs, but more recently diffusion-based models and Transformers have pushed quality further. WaveGrad~\cite{DBLP:conf/iclr/ChenZZWNC21} and DiffWave~\cite{DBLP:conf/iclr/KongPHZC21} applied diffusion processes to speech generation, achieving very high fidelity with more stable training than GANs. Transformer models like AudioLM~\cite{DBLP:journals/taslp/BorsosMVKPSRTGTZ23} have shown how to generate coherent speech or music by learning discrete audio representations and their sequences, without text transcripts. These innovations enable new communication modalities, for example, ultra-low bandwidth voice transmission by sending text or discrete codes and synthesizing speech at the receiver.

\subsection{Super-Resolution and Upscaling}

Super-resolution (SR) refers to enhancing the resolution or quality of a signal from a lower resolution version. In communication systems, SR serves as a powerful tool to save bandwidth: a low-resolution (or low-bitrate) stream is transmitted, then upscaled at the endpoint to approximate high-resolution quality.

\subsubsection{Image Super-Resolution}
Image SR has brought both quantitative and perceptual improvements in recent years. Early deep models (e.g. SRCNN~\cite{DBLP:conf/eccv/DongLHT14}, ESPCN~\cite{DBLP:conf/cvpr/ShiCHTABRW16}) optimized for PSNR, yielding high peak signal-to-noise ratio but sometimes lacking texture realism. The introduction of adversarial losses changed this: SRGAN~\cite{DBLP:conf/cvpr/LedigTHCCAATTWS17} and ESRGAN~\cite{DBLP:conf/eccv/WangYWGLDQL18} demonstrated that GANs can add realistic details (like sharp edges or textures) that make upscaled images subjectively convincing. However, GAN-based SR can introduce hallucinations, so a balance between fidelity and perceptual quality is needed. Transformer and Diffusion architectures have pushed SISR further. Vision transformers have been adapted for SR with great success. For example, SwinIR~\cite{DBLP:conf/iccvw/LiangCSZGT21} uses a Swin Transformer backbone to model long-range pixel dependencies. Similarly, Restormer~\cite{DBLP:conf/cvpr/ZamirA0HK022} introduced an efficient transformer for image restoration that set new SOTA on tasks including super-resolution, while being memory-efficient. Diffusion models, with their probabilistic refinement process, have been applied to super-resolution as well. SR3~\cite{DBLP:journals/pami/SahariaHCSFN23} is a super-resolution diffusion model that iteratively refines an image from pure noise conditioned on a low-res input, eventually producing high-res outputs.

\subsubsection{Video Super-Resolution}
Video SR builds on Image SR but leverages temporal information from neighboring frames. The past years have brought dramatic progress in video SR, thanks to advanced propagation and alignment mechanisms in deep models. Traditional video streaming could benefit greatly from video SR: a low-res video can be transmitted, and a neural video SR model at the client reconstructs it to HD or 4K. Modern video SR networks often adopt a recurrent or iterative refinement approach rather than processing each frame independently. BasicVSR~\cite{DBLP:conf/cvpr/ChanWYDL21} introduced a simple yet effective recurrent framework that propagates features forward and backward through the video clip, greatly improving detail consistency. This was soon enhanced by BasicVSR++~\cite{DBLP:conf/cvpr/ChanZXL22a}, which added second-order propagation and flow-guided deformable alignment. Another notable approach is Transformer-based video SR: while naively applying transformers to video is costly, hybrids like TTVSR~\cite{DBLP:conf/cvpr/Liu0FQ22} use a transformer for temporal fusion on tokens, and CNNs for spatial upscaling, to capture motion cues effectively.

\subsubsection{Audio Super-Resolution}
Audio SR is the task of reconstructing high-fidelity audio from a downsampled signal. Deep generative models have outperformed traditional signal processing methods in recent years. GANs, starting with SEGAN~\cite{DBLP:conf/interspeech/PascualBS17}, were early successful models for adding high-frequency components to speech. Recent advancements focus on improving fidelity and efficiency. MetricGAN~\cite{DBLP:conf/icml/FuLTL19} optimizes the generator based on perceptual metrics, enhancing quality. Diffusion models, like Universal Speech Enhancement~\cite{DBLP:conf/interspeech/ScheiblerFSK24}, gradually inject missing frequencies into a spectrogram or waveform. Flow-based models, such as WaveGlow~\cite{DBLP:conf/icassp/PrengerVC19}, generate high-resolution audio in one step, bypassing iterative sampling while modeling plausible high-frequency content.
Enhanced speech bandwidth boosts intelligibility and user experience in VoIP calls. Modern codecs include bandwidth extension, and deep learning now improves these tools' quality. These techniques are increasingly integrated into real systems.

\subsection{Quality Enhancement and Restoration}
Generative AI not only creates new content or upscale resolution but also restores and enhances degraded multimedia contents, such as images with compression artifacts, videos affected by packet loss, or audio with noise.

\subsubsection{Image and Video Artifact Removal}
Lossy compression of images and videos introduces artifacts such as blockiness, ringing, blurriness, and banding. Removing these artifacts is important for improving visual Quality of Experience on the user side. In recent years, generative adversarial approaches have proven especially effective for artifact removal, as they can synthesize missing high-frequency details rather than just smoothing them. DACAR~\cite{DBLP:conf/iccv/GalteriSBB17} showed GANs producing more photorealistic restoration of heavily compressed images than MSE/PSNR-driven methods. Building on that, multiple papers introduced enhanced networks to tackle compression artifacts. For instance, DMCNN~\cite{DBLP:conf/icip/ZhangYH018} used dual-domain (DCT and pixel domain) learning to better undo JPEG compression, and Uformer~\cite{DBLP:conf/cvpr/WangCBZLL22} applied a transformer-based architecture for image deblocking with excellent results. In video compression, research has gone into in-loop filters powered by neural networks. The latest video coding standard H.266/VVC even allows the possibility of CNN-based in-loop filtering to replace traditional filters~\cite{DBLP:conf/icip/ZhangJLL23}. Such methods are typically trained on codec-distorted frames to output cleaner versions, effectively learning the inverse mapping of the compression. In live streaming, if the decoder has GPU resources, it can apply a similar deep post-processing to every frame to improve quality without increasing the bitrate. A particularly advanced example is the use of diffusion models for video restoration. DiQP~\cite{DBLP:conf/wacv/DehaghiRM25} is a diffusion+Transformer model aimed at reversing heavy compression damage in 4K–8K video. By modeling compression artifacts as a form of noise, the diffusion process learns to iteratively denoise compressed frames, while a Transformer component captures long-range spatial-temporal dependencies. This underscores the trend: as generative models become more expressive and aware of data distribution, they can better distinguish artifacts from signals and fill in what compression removed.

\subsubsection{Denoising and Deblurring}
Denoising, the removal of random noise from images or audio, is another area that has been revolutionized. While not always a result of transmission, noise often creeps in via sensors or analog communications, and denoising is critical for clarity. Classic filters have given way to deep denoisers like DnCNN~\cite{DBLP:journals/tip/ZhangZCM017} and FFDNet~\cite{DBLP:journals/tip/ZhangZZ18}. In the last few years, as with SR, transformers and diffusion models have set new records in denoising performance. The previously mentioned Restormer~\cite{DBLP:conf/cvpr/ZamirA0HK022} not only addresses SR but also achieves SOTA in image denoising, leveraging self-attention to handle spatially varying noise effectively. It outperforms earlier CNNs and even specialized designs, especially on high-resolution images where modeling long-range correlations helps differentiate noise from signal. Researchers have applied pretrained diffusion models~\cite{DBLP:conf/icml/KulikovYKM23} to image denoising by simply using the diffusion reverse process on a noisy image conditional on a guidance signal; this has proven effective even for severe noise. Moreover, unsupervised approaches like deep image prior~\cite{DBLP:conf/cvpr/UlyanovVL18} showed that a network can be optimized to a single noisy image, implicitly modeling the clean image. For video, deep video denoising methods like DVDnet~\cite{DBLP:conf/icip/TassanoDV19} and FastDVDnet~\cite{DBLP:conf/cvpr/TassanoDV20} use multi-frame information to reduce noise while preserving motion details. As a result, it’s now feasible to clean up grainy, low-light video in real-time applications, improving visual quality for end users. In audio, speech denoising and dereverberation have also embraced GANs and diffusion models. The latest speech enhancement diffusion models can remove complex noise patterns and reverberation while preserving speech intelligibility, a task where older spectral subtraction methods struggled. These improvements directly impact VoIP and video conferencing quality by making voices clearer under adverse conditions.

\subsubsection{Error Concealment and Inpainting}
When data packets are lost in transmission (common in unreliable networks or real-time streaming), the receiver may get missing pieces of audio or video. Generative models have been applied to conceal these losses by inpainting the missing content in a plausible way. In video, traditional error concealment uses motion extrapolation from previous frames, but this often yields visible discontinuities. Recent approaches like VECGAN~\cite{DBLP:conf/eccv/DalvaAD22} employ GANs conditioned on neighboring frame content to hallucinate the lost frame regions with surprising consistency. In audio, models such as TMGAN-PLC~\cite{DBLP:conf/interspeech/GuanYLZW22} use a temporal memory GAN to generate missing speech segments from surrounding context. These models are often trained on large speech datasets with random dropouts, learning to predict plausible continuations of the waveform. In image-based communication (like wireless image transmission or live screen sharing), if parts of an image are missing or corrupted, image inpainting models can fill in the gaps. Modern inpainting GANs or diffusion-based inpainting have no trouble synthesizing content for holes even when large portions of an image are lost. Though primarily developed for photo editing, these can be repurposed for transmission errors, for example, at the decoder side of a progressive image transmission, if later packets don’t arrive, a generative model could fill the missing blocks based on context. We are beginning to see hybrid schemes: a receiver might accept a very low-quality video in bad network conditions and rely on a generative enhancement model to keep it watchable, rather than pausing playback to rebuffer. This ties into the idea of graceful degradation using AI, rather than freezing or showing blocky video, the system delivers something slightly blurry which a neural enhancer polishes in real-time. Such concepts are in early stages, but research results are promising.

\section{An Information-Theoretic Framework}

\subsection{Classical Information Theory Background}

To set the stage, we recall key principles from classical information theory~\cite{shannon1948mathematical} as a baseline. Shannon's information theory formalized fundamental concepts such as \textit{entropy} (the average uncertainty of a source), \textit{channel capacity} (the maximum reliable communication rate), and the \textit{rate-distortion function} (the lowest achievable compression rate for a given distortion tolerance). These measures, however, operate solely at the syntactic level, treating data as sequences of symbols or bits without direct consideration of their semantic content.

In classical information theory, the entropy \(H(X)\) of a discrete random variable \(X\) is defined as
\begin{equation}
H(X) = -\sum_{x\in\mathcal{X}} p(x) \log p(x),
\end{equation}
which quantifies the average uncertainty or information content of the source. This metric sets the fundamental limit for lossless compression, yet it treats all deviations uniformly, potentially overlooking variations in semantic importance.

Formally, the \textit{rate-distortion function} \( R(D) \) quantifies the minimum number of bits per symbol needed for reconstructing a source within an average distortion level \( D \). It is defined via a constrained optimization problem over all encoding schemes satisfying the distortion constraint:
\begin{equation}
    R(D) = \min_{p(\hat{x}|x):\, E[d(x,\hat{x})]\le D} I(X;\hat{X}),
\end{equation}
where \( d(x,\hat{x}) \) represents a distortion measure (for example, MSE) and \( I(X;\hat{X}) \) denotes the mutual information between the source \( X \) and its reconstruction \( \hat{X} \).

Similarly, Shannon’s \textit{channel capacity} \( C \) is the maximum mutual information achievable between the channel input and output (measured in bits per channel use) by optimizing the input distribution. These classical definitions rely on fidelity criteria such as pixel error rates or bit errors, disregarding whether such errors significantly affect the message's semantic meaning. An error altering a background pixel may have equal weighting in \( d(x,\hat{x}) \) as an error affecting a critical object, despite the latter having a far greater semantic impact.

\subsection{Towards Generative Information Theory}

In this subsection, we extend classical information-theoretic concepts to generative information theory~\cite{DBLP:series/sbcs/NiuZ25} by introducing semantic entropy, semantic mutual information, semantic channel capacity, and semantic rate-distortion theory.

\subsubsection{Semantic Entropy and Mutual Information}
Classical entropy \( H(X) \) quantifies the average surprise (in bits) associated with a random variable \( X \). Transitioning to semantics involves redefining the random variable of interest from syntactic messages to semantic representations. Formally, generative information theory typically introduces a pair of random variables \((U,\tilde{U})\), where \(U\) represents the syntactic message (e.g., a video frame) and \(\tilde{U}\) denotes the semantic content or label underlying that message. A \textit{synonymous mapping} \( f: U \to \tilde{U} \) clusters together all messages \( u \in U \) that share the same semantic meaning \(\tilde{u}\). For instance, \(U\) could index all possible video chunks, while \(\tilde{U}\) might denote the scene category or the set of objects depicted.

The \textit{semantic entropy} \(H_s(\tilde{U})\) is then defined by summing the probabilities across semantic classes:
\begin{equation}
H_s(\tilde{U}) = -\sum_{\text{semantic class } i} P(\tilde{U}=i)\log P(\tilde{U}=i),
\end{equation}
which expands explicitly in terms of the original source distribution:
\begin{equation}
H_s(\tilde{U}) = -\sum_{i}\left(\sum_{u \in U_i^s} P(u)\right)\log\left(\sum_{u \in U_i^s}P(u)\right),
\end{equation}
where \(U_i^s\) denotes the set of syntactic messages corresponding to the \(i\)-th semantic class. This quantity, measured in "semantic bits", captures the inherent uncertainty regarding message meaning, rather than the full syntactic uncertainty. Consequently, if many syntactic messages map to relatively few meanings, we have \(H_s(\tilde{U}) \ll H(U)\). Intuitively, semantic entropy sets a lower bound on the achievable compression without losing meaning. For example, multiple pixel-level variations of frames all depicting "a car stopped at a red light" can be compressed semantically into essentially the same description. Post-encoding, entropy thus reflects only the distribution of semantic scenarios rather than detailed pixel-level variability.

Extending beyond entropy, we define the \textit{semantic mutual information} \(I_s(\tilde{X};\tilde{Y})\) between the semantic content of transmitted and received signals. Consider a communication system in which the transmitter sends syntactic message \(X\) and the receiver obtains syntactic message \(Y\), with corresponding semantic variables \(\tilde{X}\) and \(\tilde{Y}\). Semantic mutual information can be defined analogously to Shannon's mutual information \( I(X;Y)=H(X)-H(X|Y) \), but now utilizing semantic entropy. Specifically, one common formulation ("up semantic mutual information") is:
\begin{equation}
I_s(\tilde{X};\tilde{Y}) = H_s(\tilde{X}) + H_s(\tilde{Y}) - H_s(\tilde{X},\tilde{Y}),
\end{equation}
where \(H_s(\tilde{X},\tilde{Y})\) represents the joint semantic entropy of input-output meaning pairs. If the communication successfully preserves meaning, \(\tilde{Y}\approx\tilde{X}\), making \(I_s(\tilde{X};\tilde{Y})\approx H_s(\tilde{X})\). Thus, semantic mutual information quantifies how many semantic bits of information the receiver gains about the transmitted message meaning. Importantly, semantic mutual information can surpass the classical mutual information \(I(X;Y)\). Even if certain syntactic bits are corrupted, the intended meaning may still be preserved, leading to:
\begin{equation}
I(X;Y) \leq I_s(\tilde{X};\tilde{Y}).
\end{equation}

Errors altering the syntactic form \(Y\) without changing semantic content \(\tilde{Y}\) reduce classical mutual information but leave semantic mutual information unaffected. This phenomenon introduces a critical concept: \textit{semantic error resilience}. A carefully designed generative communication system may intentionally tolerate certain syntactic bit-level errors or employ redundant encodings, different syntactic codewords representing identical semantic meanings, to prioritize semantic fidelity over strict syntactic correctness. Semantic mutual information then becomes a measure of effective information rate expressed through conveyed meanings.

In the context of a generative AI-based multimedia communication framework, we can interpret \(\tilde{X}\) as the ground-truth semantic segmentation or object labels and \(\tilde{Y}\) as the receiver's decoded segmentation output. The system optimization seeks to maximize semantic mutual information \(I_s(\tilde{X};\tilde{Y})\), effectively ensuring high semantic fidelity.

\subsubsection{Semantic Channel Capacity}

Semantic channel capacity (\(C_s\)) represents the maximum semantic information rate reliably conveyed over a given channel. Formally, for a discrete memoryless channel characterized by input \(X\) and output \(Y\) with associated semantic variables \(\tilde{X},\tilde{Y}\), semantic capacity is defined as:
\begin{equation}
C_s = \max_{p(x)} I_s(\tilde{X};\tilde{Y}),
\end{equation}
maximizing as usual over all possible input distributions. This mirrors the classical definition \(C = \max_{p(x)} I(X;Y)\), but utilizes semantic mutual information. Since in general \(I_s(\tilde{X};\tilde{Y}) \geq I(X;Y)\), it follows that:
\begin{equation}
C_s \geq C.
\end{equation}

Thus, semantic capacity can exceed traditional (bit-level) channel capacity. This apparently counterintuitive result occurs because semantic mutual information disregards errors that alter syntactic bits but preserve semantic meaning. For example, a completely noisy channel (with classical \(C=0\)) can still have \(C_s>0\) if transmitter and receiver share a semantic codebook enabling meaning inference despite bit-level corruption. Practically, \(C_s\) provides a new benchmark, optimizing the meaningful throughput rather than raw bit throughput. Achieving this capacity might involve intentional redundancy, semantic-level error correction, or multiple re-encodings of the same semantic concept until meaning is successfully transmitted. In generative AI based multimedia communication, for instance, a lost object's data might still be inferred contextually by the receiver, effectively increasing semantic transmission beyond what raw bits alone would suggest.

\subsubsection{Semantic Rate-Distortion Theory}

Classical rate-distortion theory characterizes the minimal rate \(R(D)\) required to encode a source \(X\) under a given distortion level \(D\) with respect to a specific distortion. In generative multimedia communication, it is extended by incorporating semantic distortion metrics and leveraging generative priors that influence both encoding and decoding. This combined framework quantifies the minimum semantic information that must be transmitted to preserve meaning, while capitalizing on prior knowledge to further reduce the required bitrate.

\paragraph{Semantic Rate-Distortion Function}  
To capture semantic fidelity, we introduce a semantic distortion measure \(d_s(\tilde{x},\tilde{x}')\) that quantifies the difference between the original meaning \(\tilde{x}\) and the reconstructed meaning \(\tilde{x}'\). The semantic rate-distortion function \(R_s(D)\) is then defined as the minimal semantic mutual information needed between the source and its reconstruction to achieve an average semantic distortion no greater than \(D\):
\begin{equation}
R_s(D) = \min_{p(\hat{x}|x)\in \mathcal{P}_D} I_s(\tilde{X};\tilde{X}'),
\end{equation}
where \(\mathcal{P}_D\) is the set of all probabilistic encodings that yield an expected semantic distortion \(\leq D\). Expanding the definition in terms of semantic entropies gives:
\begin{equation}
R_s(D) = \min_{p(\hat{x}|x)\in \mathcal{P}_D} \Bigl[H_s(\tilde{X}) + H_s(\tilde{X}') - H_s(\tilde{X},\tilde{X}')\Bigr].
\end{equation}
Intuitively, \(R_s(D)\) represents the theoretical minimum amount of semantic information required to preserve meaning below a prescribed distortion level. Since \(H_s(\tilde{X}) \leq H(X)\), this semantic formulation may achieve substantially lower rates compared to the classical \(R(D')\) for a comparable pixel-level distortion \(D'\). For instance, in multimedia communication, a traditional codec might require several megabits per second to accurately reconstruct scenes, whereas a semantic codec that transmits object masks or coordinates can achieve high semantic fidelity at dramatically lower bitrates.

\paragraph{Incorporating Generative Model Priors}  
A generative prior, denoted by \(K\), provides both the encoder and decoder with contextual knowledge that captures typical structures or patterns in the source data. Formally, consider the source as a pair \((S,X)\), where \(S\) is the intrinsic semantic state (high-level meaning) and \(X\) represents extrinsic observations (detailed content such as video frames). Although \(S\) may not be directly observable, the prior \(K\) allows the transmitter to extract or infer \(S\) from \(X\), while the receiver uses \(K\) to reconstruct \(X\) based on \(S\). Incorporating the generative prior modifies the classical R-D problem because the encoder needs to transmit only the information that is not already predicted by \(K\). This leads to a conditional semantic rate-distortion function defined as:
\begin{equation}
R_K(D) = \min_{p(m|x)} I(X; M \mid K), \quad \text{s.t.} \quad \mathbb{E}[d_S(S,\hat{S})] \leq D,
\end{equation}
where \(M\) is the encoded message transmitted over the channel and \(\hat{S}\) is the semantic reconstruction at the receiver. Here, \(I(X;M\mid K)\) quantifies the additional information (in bits) about \(X\) that must be communicated beyond what the prior \(K\) already provides. When \(K\) is highly informative, \(I(X;M\mid K)\) can be substantially smaller than the classical mutual information, and in the limiting case, where the receiver can fully infer the semantic content from \(K\), the required rate approaches zero while semantic fidelity is maintained.

\paragraph{Semantic Fidelity versus Compression Efficiency}
Introducing a generative prior enables a novel rate-distortion trade-off that prioritizes semantic meaning. Instead of minimizing pixel-wise distortion, we impose constraints on semantic distortion \(D_S\), permitting reconstructed multimedia contents to deviate visually as long as their semantic content remains accurate. Crucially, with generative priors, encoders can aggressively compress extrinsic details \(X\), focusing instead on semantic essence \(S\). The decoder’s generative prior then reconstructs detailed appearances from minimal semantic cues. This results in significantly lower bitrates compared to classical approaches that attempt to preserve all pixel-level details. Formally, if semantic state \(S\) sufficiently captures the meaningful content, we generally have \(H(S)\ll H(X)\). Thus, compressing \(S\), possibly supplemented with minor side information for appearance, achieves substantially lower rates. The semantic rate-distortion limit, in an ideal case, becomes approximately:
\begin{equation}
R_{\text{semantic}}(D_S \approx 0) \approx H(S),
\end{equation}
which is typically much smaller than the classical rate \(R(D_X)\) needed for a correspondingly low appearance-level distortion \(D_X\). Therefore, a stronger generative prior, capable of capturing more intrinsic structure of \(X\), directly reduces the required bitrate to achieve a given semantic distortion threshold.

\paragraph{Bayesian Interpretation of Compression with Generative Priors}
Another insightful perspective arises from Bayesian coding. If encoder and decoder share a generative model \(P_{\text{model}}(X)\), they effectively agree upon a prior distribution over likely content. The transmitter then sends only posterior information regarding the actual source sequence \(X\), given this prior. Specifically, the transmitter encodes an index or latent representation \(z\), and the decoder employs the generative prior to reconstruct:
\begin{equation}
\hat{X} = G(z).
\end{equation}
If the latent representation \(z\) has substantially lower dimensionality (or entropy) than the original data \(X\), substantial compression gains result. In classical terms, the original R-D function \(R(D) = \inf I(X;\hat{X})\) (minimizing mutual information for given distortion \(D\)) is significantly reduced by leveraging the generative prior. In the ideal scenario, where the generative prior accurately reconstructs most content with negligible semantic distortion, the mutual information required approaches zero. Thus, generative priors effectively act as powerful side information, formally enhancing compression efficiency through conditional mutual information formulations within extended rate-distortion theory.

\subsection{Computing Semantic Distortion}

For the application of generative information theory, a key challenge is how to effectively compute semantic distortion beyond pixel-level distortion~\cite{Qualinet}. Based on current techniques, several promising solutions have emerged. It should be noted that these methods are not independent and can be combined for use.

\subsubsection{Feature-Based Comparison Using Pretrained Models} One common approach for assessing multimedia content similarity involves feature-based comparison using pretrained models. In this method, both the original and reconstructed multimedia content are passed through a pretrained network designed to capture high-level semantic representations. The resulting feature vectors are then compared using a suitable distance metric, such as cosine distance, which effectively measures angular differences between vectors and is particularly useful when feature magnitudes vary less than their directions. This approach offers robustness to variations in lighting, scale, and viewpoint, ensuring that semantic content remains comparable even when pixel-level details differ.

\subsubsection{Multimedia Captioning for Semantic Comparison} An alternative approach for measuring semantic distortion in multimedia content involves the use of multimedia captioning systems~\cite{DBLP:conf/cvpr/ZhouZCSX18,DBLP:journals/tetci/LiTLF19}. In this method, a captioning model generates natural language descriptions for both the original and reconstructed content. The resulting captions are then compared using natural language evaluation metrics such as BLEU~\cite{DBLP:conf/acl/PapineniRWZ02} and METEOR~\cite{DBLP:conf/acl/BanerjeeL05}, which assess the overlap in meaning between sentences. These metrics offer an interpretable and quantitative measure of semantic fidelity. Moreover, because this approach produces human-readable descriptions, it enables direct assessment of semantic consistency from a viewer’s perspective.

\subsubsection{Downstream Task Performance} A third approach to evaluating semantic distortion focuses on the performance of downstream tasks that rely on multimedia content. In this method, task-specific models, such as object detection~\cite{DBLP:journals/pieee/ZouCSGY23} or scene classification~\cite{DBLP:journals/pieee/ChengHL17}, are applied to both the original and reconstructed content. The resulting performance metrics, such as detection accuracy or classification F1-score, are then compared. If these metrics show minimal degradation, it suggests that the semantic information essential to the task has been preserved despite potential losses in low-level details. This strategy grounds semantic evaluation in practical application outcomes, aligning the assessment closely with user requirements and real-world utility.

\subsection{Modeling Generative Priors}

In generative multimedia communication systems, prior knowledge from generative model can be leveraged to reduce the amount of information that needs to be transmitted. In an information-theoretic framework, such a prior is modeled as side information available to both the transmitter and the receiver. We discuss how to model these priors and, in particular, how to measure the information contained within them.

The modeling of these priors involves the following key ideas: \textit{Learning a Prior Distribution:} A prior probability distribution \(P(x \mid K)\) is learned from a large corpus of multimedia data. This distribution captures the common patterns and structures within the data. Both the sender and receiver have access to this distribution, denoted by \(K\), which serves as a baseline for predicting the multimedia content. \textit{Incorporating the Prior into Encoding:} With the prior \(K\) available, the transmitter can focus on encoding only the information that deviates from the expected patterns. In effect, the encoder transmits the residual information that is not predicted by the prior. This approach reduces redundancy by eliminating predictable components from the transmission. \textit{Conditional Information Measures:} The effect of the prior is captured by conditional entropy and mutual information measures. Instead of the classical entropy \(H(X)\) for a source \(X\), we consider the conditional entropy \(H(X \mid K)\), which quantifies the remaining uncertainty once the prior is taken into account. Similarly, the required transmission rate can be characterized by the conditional mutual information \(I(X;M \mid K)\), where \(M\) denotes the encoded message.

A key question in modeling generative priors is how to quantify the amount of information that these models capture. Two important factors come into play: the size of the model and the scale of the training dataset. Recent work on scaling laws~\cite{DBLP:journals/corr/abs-2001-08361} has shown that model performance improves predictably as both the model size and the training data increase. These improvements can be interpreted in information-theoretic terms.

\subsubsection{Model Size and Information Capacity}  
The number of parameters in a model is often taken as a proxy for its capacity to capture complex data distributions. In an idealized scenario, one might approximate the information content of a model by the number of bits needed to describe its parameters. For instance, if a model has \(N\) parameters and each parameter is stored with a precision of \(b\) bits, a naive upper bound on the model’s description length is \(Nb\) bits. However, due to parameter redundancies and correlations, the effective information captured by the model is typically lower.

\subsubsection{Training Dataset Size and Scaling Laws}  
Empirical scaling laws indicate that as the training dataset size increases, models learn more about the underlying distribution, thereby reducing the conditional entropy \(H(X \mid K)\). Better-trained models serve as more informative priors, resulting in a more compact representation of the multimedia content. This relationship can be formalized by examining how the generalization error and the negative log-likelihood on held-out data decrease with the size of the training dataset. As these metrics improve, the model's prediction of typical multimedia content becomes more accurate, effectively lowering the residual entropy that must be transmitted.

\subsubsection{Implications for Multimedia Communication}  
By integrating these measurements into an information-theoretic framework, one can model the generative prior \(K\) not only as a static side information source but also quantify its effectiveness. For example, the conditional entropy \(H(X \mid K)\) can be seen as a function of both the model’s effective capacity and the scale of the training data:
\[
H(X \mid K) = f\bigl(\text{model capacity},\, \text{dataset size}\bigr),
\]
where \(f(\cdot)\) is a decreasing function as either the model capacity or the dataset size increases. In practice, this means that larger, better-trained models yield a lower \(H(X \mid K)\), allowing the system to transmit only the residual uncertainty \(H(X_{\text{res}})\) with fewer bits. This approach directly translates into improved compression efficiency and reduced transmission rates in multimedia communication.

In summary, modeling generative priors for multimedia communication involves learning a prior distribution from extensive data and integrating this prior into the encoding process through conditional information measures. Quantifying the information in these generative models via their size and training dataset, as guided by scaling laws, provides a framework for understanding how much redundancy can be removed from the source signal. This, in turn, leads to more efficient communication protocols that transmit only the novel, unpredictable information.

\section{Future Opportunities}

\subsection{Emerging Application Scenarios}

\subsubsection{Real-Time Conferencing and Telepresence}
Generative models will enable ultra-realistic, low-bandwidth video conferencing and holographic telepresence~\cite{DBLP:conf/mm/Zhang0Z024,DBLP:conf/mm/JinD0024}. Rather than transmitting raw high-resolution video, future systems may send only essential semantic cues (e.g. positions of facial landmarks, expressions, and motions) and reconstruct detailed visuals at the receiver~\cite{DBLP:conf/nossdav/Duan0Z025}. For example, in a 3D holographic meeting~\cite{DBLP:journals/tcsv/JinDHWL25}, the network might not need to carry every pixel of a participant’s image; instead it can transmit expressive information like facial micro-expressions and body movement, allowing the receiver’s generative model to render a lifelike presence. This semantic approach to telepresence could drastically reduce required data rates while preserving conversational realism and immersion.

\subsubsection{Immersive AR/VR Experiences}
Applications in Augmented and Virtual Reality (AR/VR) stand to benefit from generative AI-driven communication~\cite{DBLP:journals/corr/abs-2309-05658}. Interactive metaverse environments~\cite{DBLP:conf/mm/DuanLFLW021} demand the real-time exchange of rich multimedia far beyond the capacity of today’s networks~\cite{DBLP:journals/network/JinLHW24}. By leveraging generative models at the edge, a user’s device can locally synthesize high-fidelity scenes or objects, guided by concise semantic descriptions from the sender. For instance, one could transmit a latent representation or a compact prompt for a virtual scene and allow a diffusion model at the receiver to generate the full detailed environment. Such a paradigm offloads intensive content creation to generative models, alleviating the data demands of AR/VR and ensuring low-latency, immersive experiences even under constrained bandwidth.

\subsubsection{Low-Resource and Remote Connectivity}
In regions with limited infrastructure or during network outages~\cite{DBLP:conf/nossdav/ZhaoF0L23}, generative AI offers a pathway to maintain communication services~\cite{DBLP:journals/comsur/ZhouHYCJCWYJWLZWL25}. By deploying lightweight models on devices, only minimal high-level information needs to be sent, and missing details can be predicted or filled in by the model. For example, an edge device might locally predict what a sender is conveying (within acceptable uncertainty) and generate the content without requiring a full data stream. This approach, essentially “replacing communication with prediction,” can keep services running when bandwidth is scarce. It also pairs well with disaster response and low-power IoT scenarios, edge devices equipped with generative capabilities can operate autonomously when cloud connectivity is unreliable. Overall, generative communication enriched by AI generation holds promise for bridging the digital divide, delivering rich multimedia information in low-resource settings by sending only the most informative pieces.

\subsubsection{Human–AI Collaboration}
Beyond human-to-human communication, generative AI will support new forms of human–AI interaction in multimedia channels. Consider remote robotic control~\cite{DBLP:journals/ijrr/KellyCHHMRWZC11} or autonomous vehicles~\cite{DBLP:journals/micro/KatoTINTH15} sharing situational awareness: a generative model could summarize a complex sensor scene into a semantic description and regenerate it for a remote operator. Early studies in autonomous driving hint at these possibilities, generative communication frameworks can integrate images and text to guide vehicles, reducing data loads and improving real-time decision-making. In telepresence applications like remote surgery~\cite{DBLP:conf/miccai/WuHSCCW24} or virtual tourism~\cite{DBLP:conf/mm/OkuraKY12}, generative AI could similarly convey crucial contextual information with minimal latency, ensuring the remote experience is functionally identical to being on-site.

\subsection{Model Development and Deployment}
Implementing the above vision requires overcoming significant technical challenges. Generative models must be reimagined to fit the stringent requirements of communication systems, energy-constrained devices to real-time operation and security. We highlight key directions for model development and deployment.

\subsubsection{Lightweight Generative Models for the Edge}

The size and complexity of state-of-the-art generative models present a barrier to their deployment in distributed networks and on user devices. Future research is converging on small, efficient models that retain high generative fidelity. Techniques like knowledge distillation~\cite{DBLP:journals/ijcv/GouYMT21} and quantization~\cite{DBLP:conf/ijcnn/LiHCRJLFG24} have shown that large models can be compressed to a fraction of their size with minimal loss in quality. Such small generative models could run on smartphones, AR glasses, or edge devices~\cite{DBLP:journals/iotj/JinLWC23}, enabling local content generation without offloading everything to the cloud. Advancing this line of work involves not only model compression but also neural architecture search for simpler generative networks and leveraging modular or multi-scale models that can operate within tight memory and power budgets.

\subsubsection{Latency-Aware Training and Inference}
In communication, timeliness is critical. Even the most impressive generative model is of limited use if it cannot operate within the millisecond-level delays required for interactive multimedia~\cite{DBLP:journals/wcl/QiaoMGFXB24}. Future generative AI development will emphasize real-time performance. This includes training strategies that account for latency, for example, encouraging diffusion models to converge in fewer denoising steps or enabling transformers to generate streaming outputs progressively. It also involves system-level optimization like model pruning~\cite{DBLP:conf/iclr/LiuSZHD19} and hardware acceleration~\cite{DBLP:journals/pieee/DengLHSX20} so that inference can be done under strict delay constraints. As an illustration, running generative AI on edge devices eliminates round-trip latency to the cloud, ensuring faster responses for things like autonomous driving and live translation. Researchers are exploring anytime algorithms (models that refine outputs if time allows, but produce a useful result quickly) and pipeline parallelism to overlap communication and computation. The future goal is “latency-aware” generative models that gracefully trade off fidelity for speed, ensuring generative communications meet real-time Quality of Service demands.


\subsubsection{Adaptive and Contextual Generation}
A practical challenge for deployment is ensuring that generative models can adapt to varying network conditions and user contexts~\cite{DBLP:journals/expert/ZhouXSVY10}. For example, a model might need to switch to a lower-detail generation mode when bandwidth drops or a device’s battery is low. Future systems could implement multi-fidelity generative coding, where the transmitter and receiver negotiate the semantic detail level based on current channel conditions (e.g. a coarse sketch versus a photorealistic image, depending on what can be supported). This requires training models that can condition on bitrate or latency targets and still produce meaningful outputs. Another direction is online learning~\cite{DBLP:journals/ijon/HoiSLZ21} and personalization: generative models that continuously learn from the user’s data~\cite{DBLP:conf/mm/JinL0C22,DBLP:conf/mm/HuYJLCZ023} to better match individual preferences or the specific semantics relevant to that user. Lightweight fine-tuning~\cite{DBLP:conf/aaai/ZhaoHHWMZJWAWZC25} or federated learning schemes~\cite{DBLP:journals/kbs/ZhangXBYLG21} might allow on-device generative models to improve over time without central retraining. Such adaptability will make generative communication systems more resilient and personalized, aligning with the semantic goal of sending what matters most to each situation.

\subsection{Cross-Modal Communication}
Future communication systems will increasingly handle multiple modalities simultaneously, such as video, audio, images, text, and even haptic or sensory data. Generative models, with their cross-modal capabilities, are poised to become the glue that binds these modalities into a unified semantic pipeline.

\subsubsection{Joint Modeling of Multimedia Content}
Generative AI provides a common representation space for disparate modalities. Modern multimodal models~\cite{DBLP:conf/bigdataconf/WuGCWY23} can take both visual and textual inputs and generate rich outputs that mix modalities, and recent large models (e.g. GPT-4o~\cite{DBLP:journals/corr/abs-2410-21276}) demonstrate the ability to process images and text together, effectively blurring the boundaries between language and vision domains. This trend suggests that a single learned representation (a latent vector or sequence) could encode information that a decoder can realize as, say, both an image and its accompanying audio. Future research will explore unified semantic embeddings that compress a video’s visuals and soundtrack, or a slideshow’s images and narration, in a joint manner rather than separately. By capturing cross-modal correlations, for example, the way lip movements in a video align with spoken words in audio, such approaches can eliminate redundant information and improve overall compression efficiency. A cross-modal communication framework would send one stream of semantic features that suffice for reconstructing all modalities together on the receiver.

\subsubsection{Cross-Modal Generation and Recovery}
Alongside joint encoding, generative techniques enable cross-modal recovery, the ability to infer one modality from another. For instance, if a communication system drops the video stream but retains the audio, a generative model could synthesize plausible video frames synced to the audio (lip-syncing a talking head or animating a scene). Conversely, silent security camera footage could be filled with audio effects by an AI that understands the scene. While such capabilities are nascent, they are under active exploration. Recent work on sounding video generation~\cite{DBLP:conf/aaai/JeongKCL25} (generating coherent audio-track given a video, or vice versa) indicates progress in aligning modalities: researchers have begun to integrate separate audio and video diffusion models to jointly generate synchronized audiovisual content. These advances hint at future communication systems where if one modality is missing or severely compressed, the gap can be filled by AI, improving robustness and user experience. Of course, achieving seamless cross-modal generation is challenging due to the heterogeneity of data and the need for temporal alignment, but ongoing improvements in model architecture and training are steadily pushing the frontier.

\subsubsection{Multi-Modal Semantic Compression}
We foresee specialized source coding techniques that leverage generative models for multi-modal data compression. One concept is \textit{modalities as side information}: e.g., compressing audio knowing that the receiver also receives the corresponding video, and using a generative model to exploit the relationship between them. The information-theoretic underpinnings for this exist in multi-view and multi-source coding, but generative AI will provide practical algorithms to realize it. Imagine a scenario of an AR/VR telepresence where visual, auditory, and even tactile data are transmitted, rather than compress each independently, the system could transmit a core semantic description (like a high-level model of the 3D environment and events in it). The receiver’s generative engines would then render the visuals, synthesize the sounds, and perhaps trigger haptic feedback, all consistent with that shared semantic model. This aligns with the generative communication principle of sending meaning instead of raw data, now extended across modalities. Achieving this will require innovations in synchronizing modalities and ensuring fidelity in each sense, but it promises a leap in efficiency for immersive communications. Cross-modal communication research is thus a key part of the future, bringing us closer to networks that convey entire experiences rather than isolated media streams.

\section{Conclusion}
This work has proposed a semantic-aware, generative information-theoretic framework that reimagines multimedia communication for the era of generative AI. By reframing classical information-theoretic constructs through a semantic lens, we shift the optimization target from syntactic precision to human-perceived meaning. This paradigm enables communication systems to leverage generative priors to achieve high semantic fidelity at lower bitrates.

Recent advances in generative models for multimedia generation, super-resolution, and restoration already demonstrate their performance in practice. Integrating these capabilities into a principled information-theoretic framework can enable ultra-efficient and adaptive communication, particularly in bandwidth-limited settings. Future directions include lightweight, low-latency models for edge deployment, adaptive semantic coding responsive to network and device conditions, and unified cross-modal representations that convey complete experiences. Generative AI is poised to reshape multimedia transmission and redefine digital communication itself.

\begin{acks}
Part of this research was supported by an NSERC Discovery Grant.
\end{acks}

\clearpage

\bibliographystyle{ACM-Reference-Format}
\bibliography{reference}

\end{document}